\documentclass{iau} 
\usepackage{graphicx}
\title[Brownian motion of SMBH]{Brownian motion of supermassive black holes in galaxy cores}
\author[P. Di Cintio, L. Ciotti  \& C. Nipoti]{Pierfrancesco Di Cintio$^1$, Luca Ciotti$^2$ \and Carlo Nipoti$^2$}
\affiliation{$^1$IFAC-CNR, Via Madonna del piano 10,
I-50019, Sesto Fiorentino (FI), Italy \\ email: {\tt p.dicintio@ifac.cnr.it} \\[\affilskip]
$^2$Department of Physics and Astronomy, Bologna University, \\ Via Piero Gobetti 93/2
I-40129 Bologna, Italy}
\pubyear{2019}
\volume{351}  
\jname{Star Clusters: From the Milky Way to the Early Universe}
\editors{A. Bragaglia, M.B. Davies, A. Sills \& E. Vesperini, eds.}
\begin{document}
\maketitle
\begin{abstract}
We investigate the dynamics of supermassive black holes (SMBHs) in galactic cores by means of a semi-analytic model based on the Langevin equation, including dynamical friction and stochastic noise accounting for the gravitational interactions with stars. The model is validated against direct $N$-body simulations of intermediate-mass black holes in stellar clusters where a realistic number of particles is accessible. For the galactic case, we find that the SMBH experiences a Brownian-like motion with a typical displacement from the geometric center of the Galaxy of a few parsecs, for system parameters compatible with M87.
\keywords{stellar dynamics, black hole physics, methods: n-body simulations, methods: statistical.}
\end{abstract}
\firstsection 
\section{Introduction}
Here we report the preliminary results of our exploration of the dynamics of supermassive black holes (SMBHs) under the effect of gravitational scattering with the stars in galaxy cores.\\
\indent Due to the high computational cost of direct $N-$body simulations, and the difficulty to attain realistic black hole-to-star mass ratios $M_{\rm BH}/m_*$ in the latter, we adopt a stochastic differential equation approach to simulate the effect of multiple gravitational ``collisions" with field stars. At variance with previous studies involving the solution of the Fokker-Planck equation for the probability distribution function of position $\mathbf{r}$ and velocity $\mathbf{v}=\dot{\mathbf{r}}$ of the black hole (BH) [e.g. see \cite{2002ApJ...572..371C,merritt15} and references therein], here we use instead the Langevin equation 
\begin{equation}\label{langeq}
\ddot{\mathbf{r}}=-\nabla\Phi_{\rm tot}(\mathbf{r})-\eta(\mathbf{r},\mathbf{v})\mathbf{v}+\mathbf{F}(\mathbf{r}),
\end{equation}
where $\Phi_{\rm tot}$ is the smooth potential generated by the stellar and dark matter distributions, $\eta(\mathbf{r},\mathbf{v})$ is the dynamical friction [\cite{1943ApJ....97..255C,1949RvMP...21..383C}] coefficient, and $\mathbf{F}(\mathbf{r})$ a fluctuating force component (per unit mass).
\section{Methods}
We model the density profiles $\rho_*$ and $\rho_{DM}$ of stars and dark matter, respectively, using the family of spherical $\gamma-$models   
\begin{equation}\label{dehnen}
\rho(r)=\frac{\gamma-3}{4\pi}\frac{Mr_c}{r^\gamma(r+r_c)^{4-\gamma}},
\end{equation}
where $M$ is the total mass, $r_c$ the core radius, and $\gamma$ the central logarithmic density slope. In addition, for the comparisons with direct $N-$body simulations, we use the Plummer model
\begin{equation}\label{plummer}
\rho(r)=\frac{3}{4\pi}\frac{Mr_c^2}{(r_c^2+r^2)^{5/2}}.
\end{equation}
The dynamical friction coefficient evaluated for a single component model with  number density $n_*(\mathbf{x})$, particle mass $m_*$, mass density $\rho=n_*m_*$ and velocity distribution $f(v_*)$ is
\begin{equation}\label{eta}
\eta=4\pi G^2n_*m_*(M_{\rm BH}+m_*)\ln\Lambda\frac{\Psi(v)}{v^3},
\end{equation}
where $G$ is the gravitational constant, $\ln\Lambda$ is the Coulomb logarithm, $v=||\mathbf{v}||$, and
\begin{equation}\label{velfunct}
\Psi(v)=4\pi\int_0^{v}f(v_*)v_*^2{\rm d}v_*,
\end{equation}
is the fractional velocity volume function. We assume that $\eta$ depends on $\mathbf{x}$ through $n_*$ [see e.g \cite{2014ApJ...795..169A,2014MNRAS.444.3738A,2014ApJ...785...51A}] and that $f$ is approximated by a Maxwellian with velocity dispersion $\sigma(r)$ obtained solving the Jeans equations for the given density profile (\ref{dehnen}) or (\ref{plummer}). In this case 
\begin{equation}
\Psi(v)={\rm Erf}\left(\frac{v}{\sqrt{2}\sigma}\right)-\frac{2v\exp(-v^2/2\sigma^2)}{\sqrt{2\pi}\sigma},
\end{equation}
so $\Psi(\infty)=1$, \cite{bt08}.\\
\indent The norm $F$ of the stochastic acceleration term in Equation (\ref{langeq}) is sampled from the \cite{1919AnP...363..577H} distribution
\begin{equation}\label{holtsmark}
H(F)=\frac{2}{\pi F}\int_0^\infty\exp\left[-\alpha(\xi/F)^{3/2}\right]\xi\sin(\xi){\rm d}\xi;\quad \alpha=(4/15)(2\pi G m_*)^{3/2}n_*,
\end{equation}
introduced originally in the context of plasma physics, and used in stellar dynamics by \cite{1942ApJ....95..489C,1943ApJ....97....1C} to study the fluctuations of the gravitational field acting on a test star.\\
\indent As Equation (\ref{holtsmark}) cannot be written explicitly in terms of simple functions, one either solves numerically the integral, or expands in series the integrand up to the desired order integrating separately the terms of the sum [see e.g. \cite{HUMMER19861}]. In the limit of large $F$, Equation (\ref{holtsmark}) is well approximated [see \cite{1980PhR....63....1K,1986SvAL...12..237P,1999EL.....46..127G,2002EL.....57..315B}] by
\begin{equation}
\tilde H(F)\sim 2\pi n_*(Gm_*)^{3/2}F^{-5/2}.
\end{equation}
We integrate Eq.~(\ref{langeq}) with the {\it quasi-symplectic} scheme of \cite{2004PhRvE..69d1107M}, which for the one dimensional case reads
\begin{equation}\label{mannella}
v(t+\Delta t)=c_2\left[c_1v(t)+\Delta t \nabla\Phi(x^\prime)+d_1 \tilde F(x^\prime) \right]\quad x(t+\Delta t)=x^\prime+\frac{\Delta t}{2}v(t+\Delta t),
\end{equation}
where $x^\prime=x(t)+{\Delta t}/{2}v(t)$. In the equations above $\Delta t$ is the constant time-step, $\tilde F$ the adimensional stochastic force, $c_1=1-{\eta\Delta t}/{2}$, $c_2=(1+\eta\Delta t/2)^{-1}$, $d_1=\sqrt{2\tau\eta\Delta t}$ and $\tau$ is fixed by the standard deviation of the distribution of $F$, $\langle F(x,t) F(x,t_*)\rangle=2\eta\tau\delta(t-t_*)$.\\
\begin{figure}[ht!]
\begin{center}
 \includegraphics[width=4.3in]{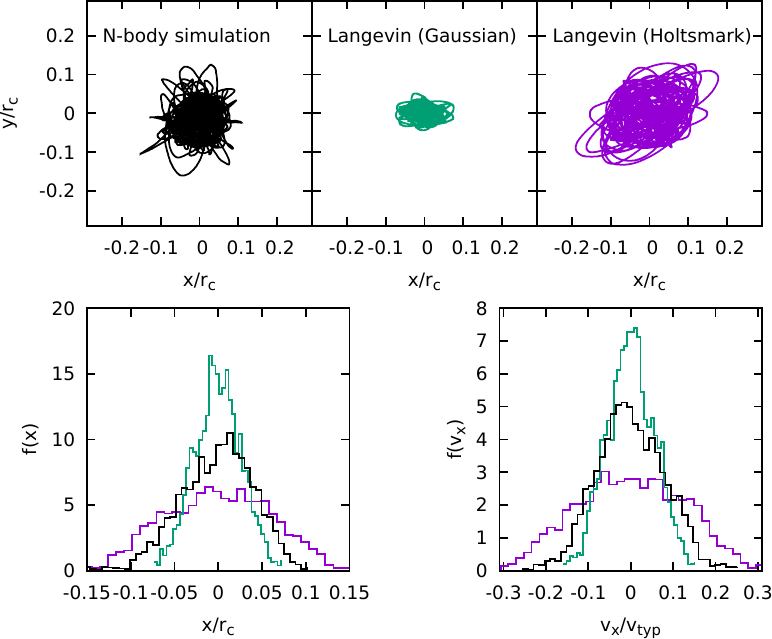} 
 \caption{Top row: orbit projections in the $(x,y)$ plane for a BH initially at rest at the centre of an isotropic Plummer model with $N=10^4$ particles in a direct $N-$body integration (left), and stochastic simulation with Gaussian (middle) and Holtsmark (right) random force distribution. In all cases $M_{\rm BH}/m_*=100$. Bottom row: distribution of positions along 
$x$ attained by the BH (left) and associated distributions of $v_x$ (right).}
   \label{fig1}
\end{center}
\end{figure}
\indent Since the Holtsmark distribution has by construction singular standard deviation, we are forced to pose a cut-off large $F$ in order to make it re-normalizable and usable in the Mannella scheme. Note that, for vanishing $\eta$ and $\tau$, Equations (\ref{mannella}) yield back the standard symplectic leapfrog method. 
\section{Results and discussion}
Equation (\ref{langeq}) can be solved with any desired mass ratio $M_{\rm BH}/m_*$ and density profile $\rho$. In order to test the validity of our model we have compared the orbit of a BH in a system of $3\times10^4$ stars distributed with a Plummer profile with mass ratio $M_{\rm BH}/m_*=100$ obtained with a standard direct $N-$body solver and with our stochastic method. In the case of the stochastic method we explored two different forms of the noise term, where $F$ is sampled from a Gaussian and a truncated Holtsmark distribution.\\
\indent In Figure \ref{fig1} (top panels) we show from left to right the orbit projections in the $(x,y)$ plane for the BH propagated in a direct $N-$body simulation and with the Langevin solver with Gaussian and Holtsmark noise terms. It appears that the case using the Holtsmak distribution better approaches the results of the $N-$body simulation (at least in term of radial displacement). In Figure \ref{fig1} (bottom panels), we show the distribution $f(x)$ of displacement along $x$ and the associated velocity component distribution $f(v_x)$, revealing that the $N-$body system has a somewhat intermediate behaviour between the stochastic models with Holtsmark and the Gaussian noise.\\
\indent Once established the convergence between our model and the direct collisional $N-$body dynamics we have investigated the possibility that certain SMBHs that appear to be off-centered in the parent galaxy 
\begin{figure}[ht!]
\begin{center}
 \includegraphics[width=4in]{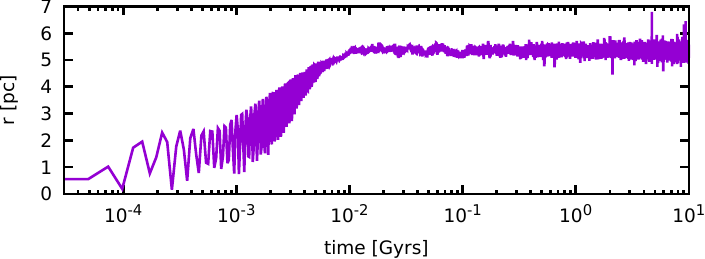} 
 \caption{Evolution of the distance from the galactic center of SMBH of $M_{\rm BH}=6\times 10^9M_{\odot}$ in a $M_{\rm gal}=3\times 10^{12}M_{\odot}$ $\gamma-$model with $\gamma=1.2$ obtained with Holtsmark noise.}
   \label{fig2}
\end{center}
\end{figure}
[see e.g. the case of M87; \cite{batch2010}; but see also \cite{geb2011}] are effectively diffusing due to several dynamical encounters with stars.\\
\indent In Figure \ref{fig2}, we show the evolution of the radial coordinate $r$ measured from the centre of the parent model (i.e. the centre of mass of the galaxy) for a $M_{\rm BH}=6\times 10^9M_{\odot}$ SMBH starting at rest in a galaxy with stellar and dark matter density distributions (2.1), total mass $M_{\rm gal}=3\times10^{12}M_{\odot}$, $\gamma= 1.2$ (for both stellar and dark components), $r_c=3$ kpc, and a dark to visible matter ratio of $\approx 6$, [parameters roughly corresponding to the case of M87, e.g. \cite{wu2006}, \cite{doi18}]. We observe that over a time of 10 Gyrs the SMBH reaches radii of the order of $\approx 6$ pc, that is compatible with the off-centre displacement claimed for the SMBH of M87, only due to multiple dynamical collisions with stars.\\
\indent We suggest that the advantages (for instance in terms of computational time) given by the stochastic models can be crucial in the study of the processes involving  SMBHs and stellar encounters. 

\end{document}